# Artificial Intelligence Generates Stereotypical Images of Scientists but Can Also Detect Them: A Pilot Study Using the Draw-A-Scientist Test

Gyeonggeon Lee*

*Natural Sciences and Science Education Department, National Institute of Education, Nanyang Technological University, 1 Nanyang Walk, Singapore 637616, Singapore*

*corresponding author: gyeonggeon.lee@nie.edu.sg

**ABSTRACT**

How the general public perceives scientists has been of interest to science educators for decades. While there can be many factors of it, the impact of recent generative artificial intelligence (AI) models is noteworthy, as these are rapidly changing how people acquire information. This report presents the pilot study examining how modern generative AI represents images of scientist using the Draw-A-Scientist Test (DAST). As a data, 1,100 images of scientist were generated using Midjourney v 6.1. One hundred of these images were analyzed by a science education scholar using a DAST scoring rubric. Using the data, the researcher went through prompt engineering to instruct gpt-4.1-mini to automatically analyze the remaining 1,000 images. The results show that generative AI represents stereotypical images of scientists, such as lab coat (97%), eyeglasses (97%), male gender (81%), and Caucasian (85%) in the 100 images analyzed by the researcher. However, gpt-4.1-mini could also detect those stereotypes in the accuracy of 79% in the same 100 images. gpt-4.1-mini also analyzed the remaining 1,000 images and found stereotypical features in the images (lab coat: 97%, eyeglasses: 95%, male gender: 82%, Caucasian: 67%). Discussions on the biases residing in today's generative AI and their implications on science education were made. The researcher plans to conduct a more comprehensive future study with an expanded methodology.

Keywords: artificial intelligence (AI), science education, draw-a-scientist test, stereotype, bias

**Introduction and Backgrounds**

The field of science education research has been interested in how learners perceive scientists, a key aspect of scientific literacy and public understanding of science. The Draw-A-Scientist (DAST) test has been used by researchers for more than 40 years, as it reveals students' thinking about science and scientists (Chambers, 1983). Sudent-drawn images of scientists has been analyzed by human researchers, using scoring rubrics such as the Draw-A-Scientist Checklist (DAST-C; Finson et al., 1995). Review studies of DAST revealed that stereotypical indicators (e.g., ethnicity, age, gender) exists across grade levels, gender, racial groups, and national borders (Finson, 2002) and remain globally consistent across time (Ferguson & Lezotte, 2020), calling for further research using DAST and providing feedback to students.

   The rapid development of artificial intelligence (AI) has impacted science education research field. First, generative AI tools like ChatGPT are changing how people acquire and consume information (Verma et al., 2023). Despite the benefits of AI for science, the pitfalls of its uncritical acceptance—e.g., bias and hallucinations—might threaten proper science education (Arantes, 2024). For example, training data and/or machine learning algorithms can impact generative AI's representation of scientists. If generative AI delivers biased images of scientist to the public around the globe, it might amplify the existing stereotypes about scientists (e.g., 'most scientists are male') to consequently hinder equitable STEM education. Therefore, it is imperative to explore the characteristics of AI-generated images of scientist.

   Second, the potential of AI to revolutionize science assessment with drawings has also been highlighted (Lee et al., 2023; Zhai et al., 2022). While computer vision models typically require much data to train them, the recent development of pre-trained vision language models (VLMs) such as GPT-4V and -4o families make broader users accessible to the technology (Lee & Zhai, 2025). Furthermore, they provide rationales behind the scoring process of images (Lee & Zhai, 2025). Also, previous studies used computer vision models to automatically assess conceptual models (Lee et al., 2023; Zhai et al., 2022), rather than more sophisticated constructs such as images of scientists. Therefore, it is timely to explore the potential of recent VLMs to automatically and precisely assess DAST drawings.

Building upon the above contemplations, the researcher suggests that image of scientists drawn by generative AI could be assessed using generative AI, based on the DAST framework, to reveal possible consequences influencing the public understanding of scientists. The researcher conducted a pilot study prior to an extensive examination. In this pilot study, the researcher report the results using Midjourney and gpt-4.1-mini.

*Research questions*

The research questions (RQs) of this pilot study are as follows:

    RQ1. Does AI generate stereotypical images of scientists?

    RQ2. Can AI detect stereotypical images of scientists?

**Methods**

*Data collection*

The image dataset was generated from Midjourney v 6.1, with the prompt "draw a scientist" (Chambers, 1983). The researcher iteratively ran sessions until the required sample size ($N = 1,100$) was met. The researcher intentionally did not adjust hyperparameters, to examine the normal products that general users might encounter.

*Scoring rubric*

The Draw-a-Scientist Checklist (DAST-C), which consists of 15 features that reflect stereotypes about scientist (Finson et al., 1995), was used as the scoring rubric for this study. The details are given in **Figure 1**.

*Human scoring*

Human scoring on the image data ($n = 100$) was conducted by the researcher, a PhD in science education, with experiences in research of AI in education including automatic scoring of student-drawn images and use of VLMs.

*Machine scoring*

Machine scoring on the image data ($N = 1,100$) was conducted using gpt-4.1-mini. Its API was iteratively called with an input image of scientist and the prompt that requests scoring the image based on the scoring rubric.

*Prompt engineering*

The prompt given to gpt-4.1-mini API is presented in Figure 1. The instruction applies role prompting method (Shanahan et al., 2023), assigning the machine a role as "a unbiased science education researcher that score a response to the draw-a-scientist test (DAST)". After that, it requires analyzing an input image according to <<<INSTRUCTIONS FOR USE OF THE DAST-CHECKLIST>>> and <<<DRAW-A-SCIENTIST CHECKLIST>>> (Finson et al., 1995)

> **Prompt**
>
> You are a unbiased science education researcher that score a response to the draw-a-scientist test (DAST). You will be given a drawing of scientist. Please score the drawing based on the <<<INSTRUCTIONS FOR USE OF THE DAST-CHECKLIST>>> and <<<DRAW-A-SCIENTIST CHECKLIST>>>. Your answer will be a python list of <<<(a)>>> rationale for each component in the checklist <<<(b)>>> 15 consecutive binary scores in a python list (0/1). Strictly follow the return structure as "[[<<<(a)>>>],[<<<(b)>>>]]".
>
> <<<INSTRUCTIONS FOR USE OF THE DAST-CHECKLIST>>>
> 1. Use one checklist sheet per subject drawing.
> 2. Place a maximum of only one mark per blank on the checklist.
> 3. If multiple images are present in the drawing (such as two or more scientific instruments), count the drawing as having ONE, NOT TWO.
> 4. If multiple images of persons appear, such as a group of scientists, mark for any and all stereotypical images that are present. (Note: If one member of the group is male and one is female, record a mark in the "male gender" blank but note the presence of the female in item #16.)
> 5. For item #4, the sub-category blanks for "size" are not counted into the checklist score. These subcategorical items are for more detailed analysis of the images and do not in and of themselves represent stereotypical images. The same applies to subcategory items #4b and #6a.
> 6. Any relevant captions (item #7) should be written down in item #16 if room doesn't allow under #7 iteslf.
> 7. TOTALS:
> A. Add the number of marks for the upper checklist and record in the box at the checklist's lower left corner. REMEMBER to add ONLY ONE mark per blank.
> B. Add the number of marks for the lower checklist and record in the box at the checklist's lower left corner. REMEMBER to add ONLY ONE mark per blank.
> C. Add the upper and lower checklist scores for the TOTAL SCORE, and record on the blank at the bottom right corner.
> 8. ANALYSIS:
> A. You may select to use the TOTAL SCORE for analysis purposes or you may select to use only the upper (or lower) checklist scores. Use of TOTAL SCORE provides for more variance than use of only half scores.
> B. The checklist has been used in a pretest-posttest format with ANCOVA procedures. Other analytical procedures will likely work as well.
> C. Report other data (such as those in item #16) as percentages of drawings possessing specific images.
>
> <<<DRAW-A-SCIENTIST CHECKLIST>>>
> 1. Lab Coat (usually but not necessarily white) .................. (0 / 1)
> 2. Eyeglasses .......................................... (0 / 1)
> 3. Facial Growth of Hair (beards, mustaches, abnormally long sideburns) . . . (0 / 1)
> 4. Symbols of Research (scientific instruments, lab equipment of any kind) . (0 / 1)
> 5. Symbols of Knowledge (principally books, filing cabinets, clipboards, pens in pockets, etc.) ............................. (0 / 1)
> 6. Technology (the "products" of science) ...................... (0 / 1)
> 7. Relevant Captions (formulae, taxonomic classification, the "eureka!" syndrome) .............................. (0 / 1)
> 8. Male Gender .......................................... (0 / 1)
> 9. Caucasian ........................................... (0 / 1)
> 10. Indications of Danger .................................... (0 / 1)
> 11. Presence of Light Bulbs ................................. (0 / 1)
> 12. Mythic Stereotypes (Frankenstein creatures. Jekyll/Hyde figures, "Mad/Crazed") ...................... (0 / 1)
> 13. Indications of Secrecy (signs or warnings of "Private,""Keep Out," "Do Not Enter.""Go Away.""Top Secret," etc.) ................ (0 / 1)
> 14. Scientist Doing Work Indoors ............................. (0 / 1)
> 15. Middle Aged or Elderly Scientist ........................... (0 / 1)
> NOTE: Several indicators of the same type in a single drawing count as ONE indicator (eg. Two scientists each with eyeglasses counts as one, not two).

**Figure 1**. The prompt given to gpt-4.1-mini API to analyze each draw-a-scientist test response from Midjourney (reorganized from Finson et al., 1995)

**Results**

*Images of scientist generative by AI*

In response to the prompt given by the researcher (**Figure 1**), the generative AI successfully generated images of scientist in various appearances. The example images of scientist generated by Midjourney v 6.1 are presented in **Figure 2**.

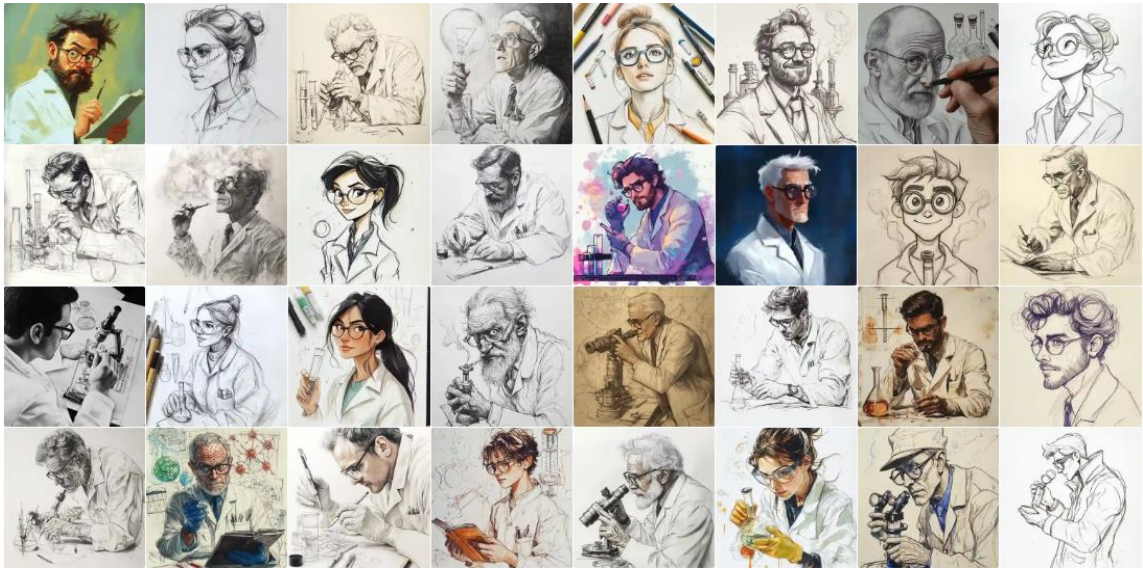

**Figure 2**. Example images generated by Midjourney v 6.1 by prompt "Draw a scientist"

*Machine-human agreement in AI-generated image scoring*

The example human and machine scorings of AI-generated image of scientist is given in Figure 3. The researcher simply coded each image for the 15 features defined in DAST-C (Finson et al., 1995). When each image was fed to gpt-4.1-mini API with the prompt (Figure 1), it returned the scoring results with its rationale for the equivalent 15 features.

For the image of scientist on the left of **Figure 3**, the MHA was 93% (14/15). The researcher and machine agreed on 14 features, except for '6. Technology (the "products" of science). While the machine thought the bulb in the figure can be considered as "a product of science, representing technology," the researcher counted it only for '11. Presence of Light Bulbs'. For the image of scientist on the right of **Figure 3**, the MHA was 93% (14/15). The researcher and machine agreed on 14 features, except for '14. Scientist Doing Work Indoors'. While the machine thought "background is not clearly indoor or outdoor," the researcher thought that experiments using round-bottom flask with personal safety equipment usually happens indoor. These show that VLMs such as gpt-4.1-mini can detect stereotypical features in the given image of scientists with rationales, based on the scoring rubric.

The results of human and machine scoring on the 100 sampled images are given in Table 1. According to human scoring, there were stereotypical features in the image of scientist, such as '1. Lab Coat' (97%), '2. Eyeglasses' (97%), '8. Male Gender' (81%), '9. Caucasian' (85%), and '5. Symbols of Research' (72%). The machine also detected similar patterns in features such as '1. Lab Coat' (98%), '2. Eyeglasses' (98%), '4. Symbols of Research' (86%), '8. Male Gender' (83%), and '9. Caucasian' (70%).

The MHA varied across the 15 features (49-100%). MHA was relatively high in concrete features such as '1. Lab Coat' (95%), '2. Eyeglasses' (95%), '10. Indications of Danger' (100%), and '13. Indications of Secrecy. (100%). In contrast, MHA was relatively low in uncertain features such as '5. Symbols of Knowledge' (58%), '14. Scientist Doing Work Indoors' (57%), and '15. Middle Aged or Elderly Scientist' (56%). The average MHA was 79%, which is fair. This shows that VLMs' scoring of image of scientist shows considerable MHA, which can be thought as a validity of the scoring machine.

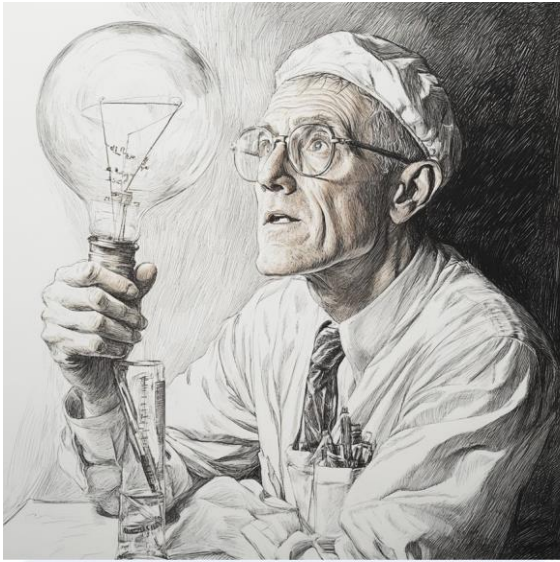
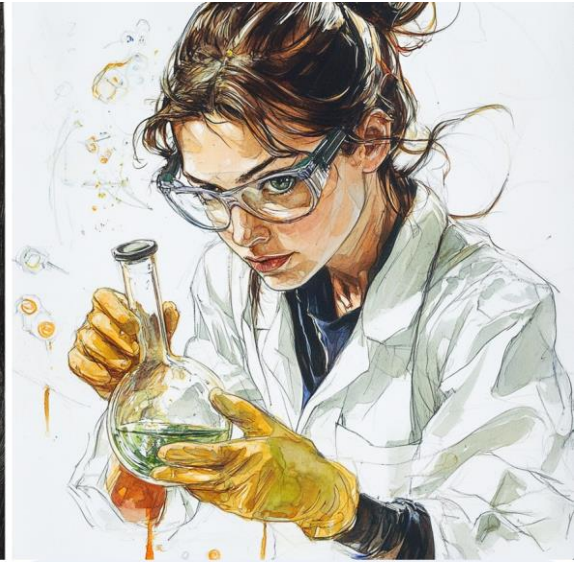

**Figure 3**. Example scoring of human and machine (gpt-4.1-mini) on images of scientist generated by Midjourney (accuracy - left: 93%; right: 93%)

**Table 1**. Results of analyses on Midjourney-generated DAST images by human and machine (%)

| Stereotypical features (0/1) | Human score (on 100 images) | Machine score (on 100 images) | Machine-human agreement | Machine score (on an additional 1,000 images) |
|---|---|---|---|---|
| 1. Lab Coat | 97% | 98% | 95% | 97% |
| 2. Eyeglasses | 97% | 98% | 95% | 95% |
| 3. Facial Growth of Hair | 52% | 49% | 49% | 51% |
| 4. Symbols of Research | 72% | 86% | 70% | 89% |
| 5. Symbols of Knowledge | 49% | 33% | 58% | 36% |
| 6. Technology | 3% | 18% | 79% | 23% |
| 7. Relevant Captions | 0% | 6% | 94% | 9% |
| 8. Male Gender | 81% | 83% | 72% | 82% |
| 9. Caucasian | 85% | 70% | 67% | 67% |
| 10. Indications of Danger | 0% | 0% | 100% | 3% |
| 11. Presence of Light Bulbs | 2% | 4% | 94% | 7% |
| 12. Mythic Stereotypes | 1% | 2% | 97% | 1% |
| 13. Indications of Secrecy | 0% | 0% | 100% | 0% |
| 14. Scientist Doing Work Indoors | 60% | 75% | 57% | 75% |
| 15. Middle Aged or Elderly Scientist | 58% | 56% | 56% | 54% |
| **Average** | - | - | 79% | - |

*Machine scoring*

The results of machine scoring on 1,000 images in addition to the abovementioned 100 images are given in **Table 1**. The scoring machine identified stereotypical features in those images, such as '1. Lab Coat' (97%), '2. Eyeglasses' (95%), '3. Male Gender' (82%), and '9. Caucasian' (67%). This shows that AI-generated images of scientist represent stereotypical perceptions of them.

**Discussions**

*AI generates stereotypical images of scientists: Implications for AI*

The results show that while generative AI models such as Midjourney can provide a user with images of scientists, those images represent stereotypical perceptions of scientists. This implicates that currently serviced generative AI models may enhance stereotypes of scientists among the public, which should be avoided. It is difficult to tell whether these stereotypes have come from - i.e., data or training algorithm. This could even be related to the current situation of STEM education. In any case, it is strongly recommended to figure out ways to correct the current representations of scientists by prominent image generation AI models.

*AI detects stereotypical images of scientists: Implications for science education*

The results show that gpt-4.1-mini, a VLM, can detect stereotypical images of scientists. This adds to the literature on automatic assessment of drawn images, scoring a more abstract construct beyond conceptual understandings. Further, the potential of AI that detect AI's bias supports using the cutting-edge technology for more efficient and still fair assessment instruments. Especially, the VLM's ability to provide rationale behind the scoring could significantly assist teachers' classroom usage of such AI tools. VLM-based automatic assessment tools might be mounted to an interactive digital learning platform, which can be used to support students in classroom settings.

*Future works*

This pilot study has several limitations in its methodology. The researcher plans to extend this pilot study to cover broader problem space on the topic. The extension includes but is not limited to the aspects summarized in **Table 2**.

**Table 2**. Plans for extended future works

| Aspects | Considerations |
|---|---|
| Image generation models | Multiple cutting-edge image generation models such as Midjourney, Stable Diffusion, OpenAI DALL·E 3, and Google Imagen3 will be considered. |
| Image generation prompt | A different prompt to generate image will be also used: "Imagine that tomorrow you are going on a trip (anywhere) to visit a scientist in a place where the scientist is working right now. Draw the scientist busy with the work this scientist does. Add a caption, which tells what this scientist might be saying to you about the work you are watching the scientist do. Do not draw yourself or your teacher." (modified DAST; Farland-Smith, 2012) |
| Sample size | More than 5,000 images of scientists will be generated. |
| Scoring rubric | Besides the DAST-C (Finson et al., 1995), a modified version of DAST scoring rubric (Farland-Smith, 2012) will also be tested. |
| Human scoring | More than or equal to two human scorers will be involved in scoring validation dataset. |
| Scoring models | Multiple cutting-edge VLMs such as OpenAI GPT-4o, Google Gemini 2.0 Flash, Meta Llama 4 Maverick, and Anthropic Claude 3.7 Sonnet will be considered. Custom computer vision models based on convolutional neural networks can be trained for comparison purpose. |
| Prompt engineering | Iterative prompt engineering will be done to increase MHA. |
| Data imbalance | The data collected in this study were quite imbalanced in terms of the 15 stereotype features (Table 1). This makes judging whether the MHA performance index (Table 1) acceptable complicated. It is true that, since pre-trained VLMs are not trained by the imbalanced data, there is no issue for overfitting. Even though, collecting more balanced data and re-doing prompt engineering will help secure validity of future studies. |

**Conclusion**

This report presents the pilot study examining how modern generative AI represents images of scientist using the Draw-A-Scientist Test (DAST). As a data, 1,100 images of scientist were generated using Midjourney v 6.1. One hundred of these images were analyzed by a science education scholar using a DAST scoring rubric. Using the data, the researcher went through prompt engineering to instruct gpt-4.1-mini to automatically analyze the remaining 1,000 images. The results show that generative AI represents stereotypical images of scientists, such as lab coat (97%), eyeglasses (97%), male gender (81%), and Caucasian (85%) in the 100 images analyzed by the researcher. However, gpt-4.1-mini could also detect those stereotypes in the accuracy of 79% in the same 100 images. gpt-4.1-mini also analyzed the remaining 1,000 images and found stereotypical features in the images (lab coat: 97%, eyeglasses: 95%, male gender: 82%, Caucasian: 67%). Discussions on the biases residing in today's generative AI and their implications on science education were made. The researcher plans to conduct a more comprehensive future study with an expanded methodology.

[Version 1 - Last modified: Apr 27, 2025]

# References

Arantes, J. (2024). Understanding intersections between genAI and pre-service teacher education: What do we need to understand about the changing face of truth in science education? *Journal of Science Education and Technology*. https://doi.org/10.1007/s10956-024-10189-7

Chambers, D. W. (1983). Stereotypic images of the scientist: The draw-a-scientist test. *Science Education, 67*(2), 255-265.

Farland-Smith, D. (2012). Development and field test of the modified draw-a-scientist test and the draw-a-scientist rubric. *School Science and Mathematics, 112*(2), 109-116.

Ferguson, S. L., & Lezotte, S. M. (2020). Exploring the state of science stereotypes: Systematic review and meta-analysis of the Draw-A-Scientist Checklist. *School Science and Mathematics, 120*(1), 55-65.

Finson, K. D. (2002). Drawing a scientist: What we do and do not know after fifty years of drawings. *School Science and Mathematics, 102*(7), 335-345.

Finson, K. D., Beaver, J. B., & Cramond, B. L. (1995). Development and field test of a checklist for the Draw-A-Scientist Test. *School Science and Mathematics, 95*(4), 195-205.

Lee, G., & Zhai, X. (2025). Realizing visual question answering for education: GPT-4V as a multimodal AI. *TechTrends. 69*, 271-287.

Lee, J., Lee, G. G., & Hong, H. G. (2023). Automated assessment of student hand drawings in free-response items on the particulate nature of matter. *Journal of Science Education and Technology, 32*(4), 549-566.

Verma, G., Campbell, T., Melville, W., & Park, B. Y. (2023). Navigating opportunities and challenges of artificial intelligence: ChatGPT and generative models in science teacher education. *Journal of Science Teacher Education, 34*(8), 793-798.

Shanahan, M., McDonell, K., & Reynolds, L. (2023). Role play with large language models. *Nature, 623*(7987), 493-498.

Zhai, X., He, P., & Krajcik, J. (2022). Applying machine learning to automatically assess scientific models. *Journal of Research in Science Teaching, 59*(10), 1765-1794.